\documentclass[prl,twocolumn]{revtex4-1} 

\usepackage{graphicx}
\usepackage{amsmath}
\usepackage{amsfonts} 
\usepackage{mathrsfs} 
\usepackage{color}
\usepackage{natbib,hyperref,url}
\usepackage{bm}

\def\<{\langle}
\def\>{\rangle}
\renewcommand{\H}{{\cal H}}

\def\beq{\begin{equation}}
\def\eeq{\end{equation}}
\def\barray{\begin{eqnarray}}
\def\earray{\end{eqnarray}}

\def\myi{\imath}

\def\bs{\beta_{\rm s}}
\def\bc{\beta_{\rm t}}
\def\db{\delta\beta}

\def\dbn{\delta\beta_{\rm d}}
\def\C{{\bm \sigma}} 
\def\f{g}

\begin{document}

\title{Anomalous metastability in a temperature-driven transition}
\author{Miguel Ib\'a\~nez Berganza$^1$,\footnote{miguel.berganza@roma1.infn.it} Pietro Coletti$^2$, Alberto Petri$^{3}$}
\affiliation{$^1$IPCF-CNR, UOS Roma {\em Kerberos} {\rm and} Dipartimento di Fisica, Universit\`a ``La Sapienza''.  Piazzale A. Moro, 5, 00185 Roma, Italy.\\ $^2$Dipartimento di Matematica e Fisica, Universit\`a Roma Tre, Via della Vasca Navale 84, 00146 Roma, Italy. \\ $^3$Istituto dei Sistemi Complessi - CNR, via del Fosso del Cavaliere 100, 00133 Rome, Italy.}

\begin{abstract}
Langer theory of metastability provides a description of the lifetime and properties of the metastable phase of the Ising model field-driven transition, describing the magnetic field-driven transition in ferromagnets and the chemical potential-driven transition of fluids.  An immediate further step is to apply it to the study of a transition driven by the temperature, as the one underwent by the two-dimensional Potts model. For this model a study based on the analytical continuation of the free energy (Meunier, Morel 2000) predicts the anomalous vanishing of the metastable temperature range in the limit of large system size, an issue that has been controversial since the eighties. With a parallel-GPU algorithm we compare the Monte Carlo dynamics with the theory, obtaining agreement and characterizing the anomalous system size dependence. We discuss the microscopic origin of these metastable phenomena, essentially different with respect to the Ising case.
\end{abstract}

\maketitle

\begin{figure}[t!]           
\begin{center} 
 \includegraphics[width=.99\columnwidth]{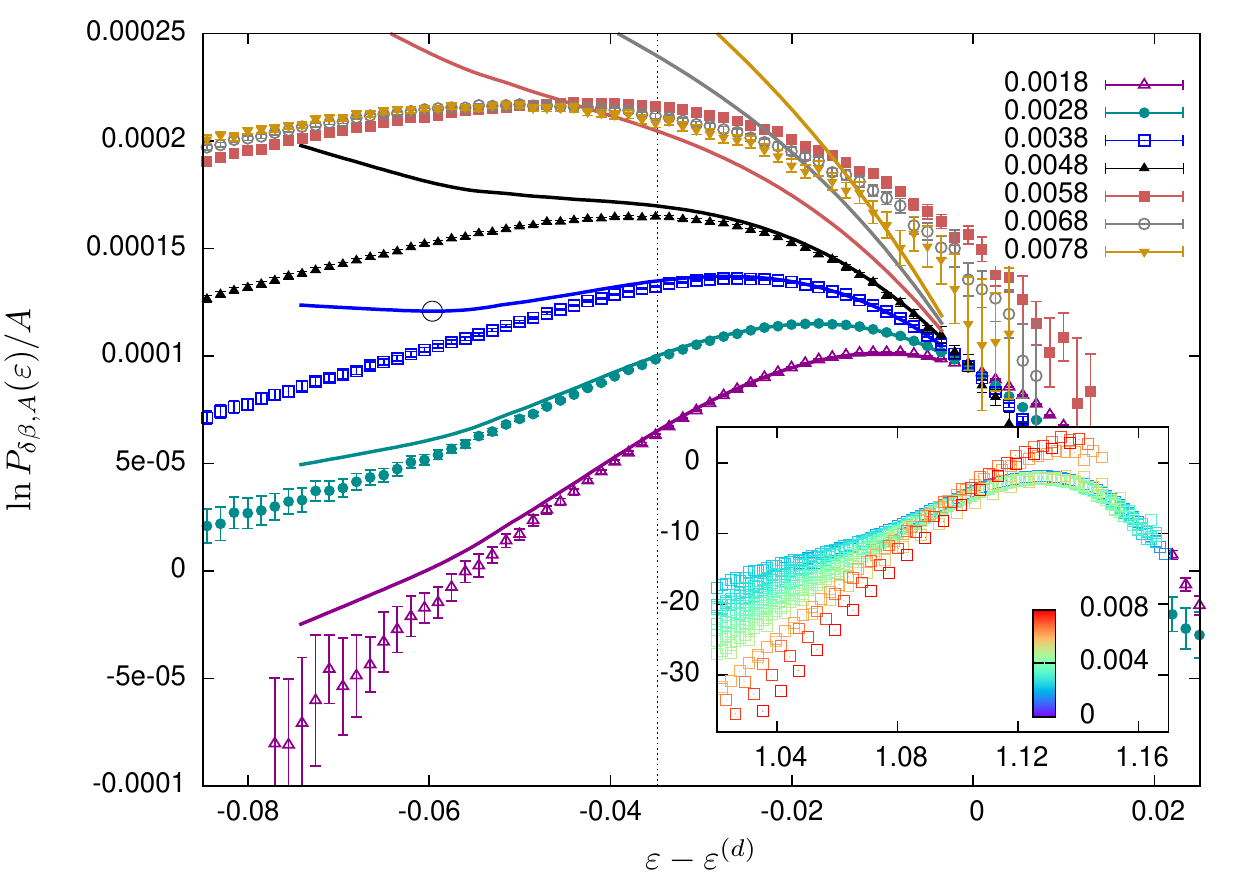} 
 \caption{Logarithm of the EPD for different inverse temperatures $\db$ (on the key) for $A=256^2$, vs. the theoretical prediction Eq. (\ref{eq:Pe}) (curves). Inset: the quantity $\ln P_{\db,A}(\varepsilon)-\db A \varepsilon$, for different $\db$s from 0.0016 to 0.08. Only for  low $\db\lesssim 0.0046$ the data overlap for different $\db$s  (they satisfy re-weighting). 
}
\label{fig:EPD}
\end{center}   
\end{figure}

\begin{figure}[t!]           
\begin{center} 
 \includegraphics[width=.9\columnwidth]{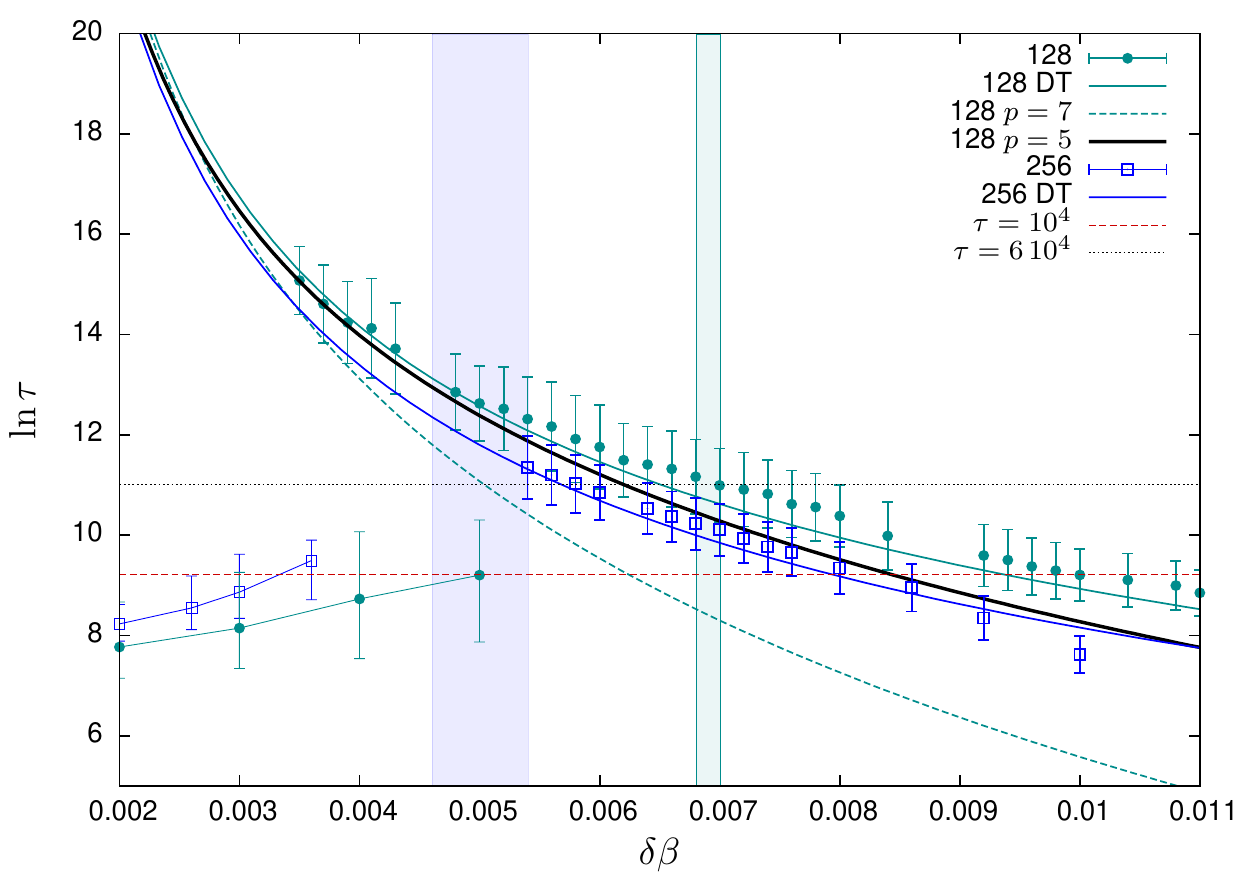} 
\caption{Logarithm of the lifetime for $A=128^2$, $256^2$, vs. the DT prediction  (full color curves) \cite{Berganza2014}, along with Eq. (\ref{eq:tauvsdb}) with $p=5$ and 7 (see text), and ${\cal K}_A$ taken as the only adjustable parameter. The vertical strips signal the error intervals around $\db^*(A)$. For small $\db$ we show $\tau_{\rm r}$, the data suggest that $\tau_{\rm r}\sim \tau$ at $\db^*$. 
}
\label{fig:tau}
\end{center}   
\end{figure}

\begin{figure}[t!] 
\begin{center} 
 \includegraphics[width=.99\columnwidth]{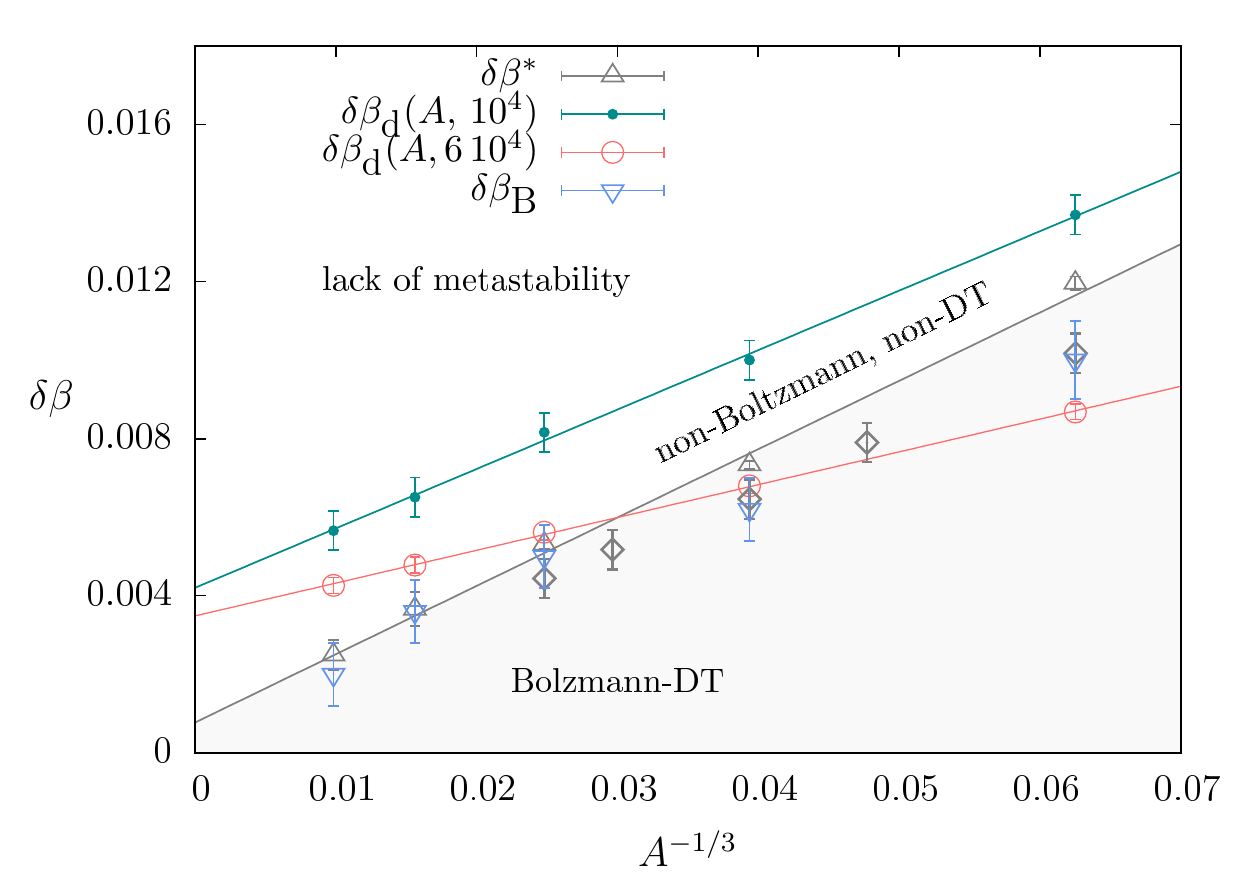} 
\caption{Pseudo-spinodal point $\db^*(A)$ extrapolated from our data vs. the DT predictions (carr\'es); $\dbn(A,t)$ for two values of $t$ and a fit with Eq. (\ref{eq:Ascaling}) (lines). We also report $\db_{\rm B}(A)$ at which the EPD stops being reweightable.  Three different dynamical finite-size regimes emerge (see main text).}
\label{fig:regimes}
\end{center}   
\end{figure}

Metastability is  ubiquitous in nature and in technology \cite{debenedetti1996metastable}. It is an important concept in many fields of physics \cite{Rikvold1995Recent,*[{see also: }]Gunter1993Numerical,*Gunther1994Application}. In particular, it is crucial in the context of the glass transition problem \cite{Debenedetti2001Supercooled,debenedetti1996metastable,Cavagna200951}. Metastability is also present in many biological systems, such as proteins \cite{Thirumalai2011protein} or nucleic acids \cite{Volker2008}. Despite the relevance of the subject, there is no general theoretical framework allowing for the computation of the lifetime and properties of the Metastable Phase (MP) provided the details of the microscopic interaction \cite{Binder1987Theory}. On the other hand, understanding how finite-size effects can influence this peculiar state of matter is becoming more and more relevant with the increasing development of miniaturization processes and of nano- and bio-sciences \cite{Uchic2004,Sandoval2009,Islam2010}.  \\
Beyond mean-field approximation, metastable states correspond to {\it local} free energy saddle points in phase space and standard methods of statistical physics are not able to capture their properties \cite{Binder1987Theory}. The main theoretical tools to tackle them are based on restricted ensembles \cite{Capocaccia1974,Penrose1971,Stillinger1995,Corti1995Metastability}, which exclude phase-separated configurations from the partition function. Of particular relevance is the study by Langer \cite{Langer1967Theory,Langer1968Theory}. He showed that the MPs of the Ising/Lattice Gas Model (ILGM) can be described by the analytical continuation of the free energy $f$ in the unstable phase, $\tilde f$. Its imaginary part, ${\rm Im }\,\tilde f$, is in this context proportional to the nucleation rate, $I$, for a wide class of model dynamics \cite{Langer1980}. Such an  approach allows for the computation of $I(h,T)$ as a function of the under-critical temperature $T$ and of the field $h>0$, corresponding to the MP with negative magnetization \cite{Langer1967Theory,Langer1980,Langer1968Theory,Gunther1980Goldstone,Binder2013Monte}. The Langer description of the ILGM has been compared with dynamical methods \cite{Rikvold1995Recent}, in which the MP is characterized as a stationarity of observables under a Markov-Chain Monte Carlo (MC) local dynamical update \cite{Binder1974Investigation,Rikvold1994Metastable,Binder1973Scaling,Heerman1984Nucleation}. A general agreement between the Droplet Theory (DT) and dynamics is found in 2D, 3D. Finite-size effects are also well understood in the ILGM \cite{Rikvold1994Metastable}, and absent when the system linear size $L$ is much larger than the length scales involved in the nucleation process.\\
While the relationship between ${\rm Im}\,f$ and $I$  has been proved for several systems, its general applicability is not known \cite{Rikvold1995Recent}. The analytical continuation of $f$ {\it \`a la} Langer has been computed for the order-disorder, $T$-driven transition of the $q$-color Potts model (PM) in 2D \cite{Meunier2000Condensation}. The interest for this model lies in the fact that, at variance with the ILGM, for $q>4$ it undergoes a discontinuous transition  driven by the temperature, which represents the unique thermodynamic variable of the model. This essential difference with respect to the ILGM motivates the application of the Langer/Fisher methods. A further motivation to study the MPs is that their existence for large system area at inverse temperatures above the inverse transition temperature $\bc$ still results unclear after a long debate. In this paper we shall investigate the effects of finite size on metastability in this paradigmatic model and settle, at least phenomenologically, the controversial about the disappearing of the MP in thermodynamic limit. \\
Binder first posed the question  \cite{Binder1981Static},  relating it to that of the finiteness or divergence of the specific heat at $\bc$, $c_{\rm t}$. This approach was recently pursued \cite{Ferrero2012} with the help of Graphic Processing Units (GPUs), resulting in a finite $c_{\rm t}$. The possible existence of a metastable interval results as well from  pseudo-critical approaches, which find critical divergences at the (so called {\it pseudo-spinodal}) temperature $\bs>\bc$ \cite{Fernandez1992,Shulke2000,Loscar2009}, for different values of $q$. 
A different picture emerge from the study \cite{Meunier2000Condensation}, the mentioned DT for the 2D $q>4$ PM. The free energy  $f_a$ of the ensemble of droplets with area $a$  is guessed, by compatibility with exact results at $\bc$, to be $\beta f_a=\db\,a - w_q a^{-2/3}-(7/3)\ln a$, being $w_q$ the surface tension energy, inversely proportional to the correlation length $\xi_q$, and $\db=\beta-\bc$ the excess of inverse temperature. The resulting free energy of the disordered (stable) phase for $\db<0$  is analytically continued for complex values of the inverse temperature, obtaining the function $\phi$: 

\barray
\phi(z) = \int _0^\infty \frac{{\rm d}\zeta}{\zeta^3}\,e^{-\zeta} \left[ e^{-z\zeta^{3/2}}-1-z\zeta^{3/2}\right]\label{eq:phi} 
\earray
where $z=-\db/w_q^{3/2}\in\mathbb{C}$. Then, the finite-size energy probability distribution  (EPD)  $P_{A,\db}(E)$ corresponding to a system with $A$ sites at $\db=0$ is obtained by a Laplace transform of $\phi$, and extended to the MP at $\db>0$ by re-weighting:  

\barray
P_{A,0}(\varepsilon) = \frac{\f A w_q^2}{2\myi \pi} \int_{-\myi\infty}^{\myi\infty}\,{\rm d}z\,e^{ w_q^2 A(\phi(z)-\epsilon z)  } \label{eq:Pe} \\
P_{A,\db}(\varepsilon)=P_{A,0}(\varepsilon)\,e^{-\db A \varepsilon} \nonumber \label{eq:reweighting} 
,
\earray
$g$ is a constant, $\varepsilon$ is the energy per site and $\epsilon=(\varepsilon-\varepsilon^{(d)})/gw_q^{1/2}$ is a rescaled energy shifted by the average (disordered phase) energy, $\varepsilon^{(d)}$, at $\bc$ \cite{Meunier2000Condensation}. $P_{A,0}$ happens to exhibit an anomalous slow dependence on the area, such that the metastable interval $[\bc:\bc+\beta^*(A)]$ in which the EPD presents a convex region shrinks to zero in the large-$A$ limit. No metastability is found to exist for $A\to\infty$. The disappearance of the metastable interval for large $A$ is also found in \cite{Berganza2007}, by means of  {\it MC dynamics} in the MP at small $\db>0$. The metastable dynamics at $\db>0$ has also been studied in \cite{Nogawa2011Static}, where an anomalous finite-size behavior has been pointed out. \\ 
\indent
There are several questions to be clarified. First, the apparent disagreement between the results of pseudo-critical attempts \cite{Fernandez1992,Shulke2000,Loscar2009}, indicating the existence of a metastable interval for large sizes, and those based on stability conditions of the MP \cite{Meunier2000Condensation}, and on dynamics \cite{Berganza2007,Nogawa2011Static}. Secondly, to what extent the results of Ref. \cite{Meunier2000Condensation} coincide with  those of a dynamic sampling for $\db >0$  \cite{Meunier2000Condensation,Nogawa2011Static,Berganza2007}. Finally, the microscopic origin of the shrinking of the metastable interval for large sizes, which is absent in the ILGM paradigm. In this letter we perform a comparison (the first one, to our knowledge) between the results from the DT and the dynamical averages under a local MC updating rule. We report evidence of the {\it dynamical} metastable inverse temperature interval shrinking approximately as $\sim A^{-1/3}$. This scaling can be also derived from the analysis of \cite{Meunier2000Condensation} in saddle point approximation. A devoted GPU parallel algorithm has been developed for the efficient simulation of large sizes ($L\sim 1024$). We have also devised a method allowing for an accurate comparison with the DT \cite{Meunier2000Condensation}, an  unaccomplished task, to our knowledge. 
\indent
Let us describe our method. The  $q$-PM \cite{Wu1982Potts} is defined in a configuration of $A$ spins $\{\sigma_j\}_{j=1}^{A}$, $\sigma_j=1,\ldots,q$, with Hamiltonian $\H=-\frac{1}{2}\sum_{i,j} \delta_{\sigma_i,\sigma_j} {\cal A}_{ij} $, where $\cal A$ is the adjacency matrix, which corresponds to a 2D lattice with periodic boundary conditions. For $A\to\infty$, the model presents a first-order phase transition for $q>4$ at an inverse temperature $\bc(q)=\ln(q^{1/2}+1)$  \cite{Wu1982Potts}. Our analysis is for $q=12$, for which $\varepsilon^{(d)}\simeq -0.8637$ and $\xi_{12}\simeq 6.54$ lattice sites \cite{Borgs1992Explicit}.  Starting from different random configurations, we perform series of Metropolis MC Markov chains which generate {\it sequences} of configurations $\{\C_t\}_t$ at different times, differing by $t_{\rm s}=128$ MC steps. For $\db>0$, {\it we compute averages from sub-sequences of stationary configurations only}. As a criterion of stationarity, we  impose that the temporal self-correlation function $C_t(t')=\C_t\cdot \C_{t+t'}$, that is the overlap between configurations at instants $t$, $t+t'$, does not differ too much with respect to $C_{t-\delta t}$, the same function $\delta t=10^3$ MC steps before, the difference required to be of the order of its fluctuations in stable equilibrium  \cite{Berganza2014}. We assume that the set of different sub-sequences of stationary configurations (with energy safely larger than the corresponding ordered equilibrium energy at $\db$) constitutes a ``stationary ensemble'' to be compared with the restricted ensemble approach implicit in the metastable EPD $P_{A,\db}$ of \cite{Meunier2000Condensation}, since the convexity of the latter is expected to induce stationarity under a local dynamics. The adoption  of such a stationary ensemble yields better results than the commonly used first-passage time method \cite{Berganza2007,Nogawa2011Static}, since it excludes  non-equilibrium realizations with over-critical growing droplets \cite{Binder1987Theory,Rikvold1995Recent} and systematically discriminates fake energy plateaus \cite{Berganza2014}. Due to the low free-energy barriers of the model for moderate values of $q$, such a method is necessary for an accurate comparison with the theory \cite{Berganza2014}. Within such a stationary ensemble we measure the EPD. On the other hand, properties of the whole stationary sequence are averaged over different instances of the sequence. An example of such an average is the average time length of the stationary sub-sequence, that will be called the {\it lifetime of the MP}, $\tau(\db,A)$. \\
\indent
In Fig. \ref{fig:EPD} we present the EPD for $A=256^2$, for different values of $\db$. Symbols are data from our simulation, while continuous lines are the prediction Eq. (\ref{eq:Pe}). We have calculated the last quantity by contour-integrating Eq. (\ref{eq:Pe}) through the steepest descent of the integral, the complex evaluation of $\phi$ requiring in turn the integration through the optimal contour in Eq.  (\ref{eq:phi}). The theoretical EPDs present a minimum at $\varepsilon_{\rm m}(\db,A)$ (evidenced as the circle over the  $\db=0.0038$ curve), which increases with $\db$. Within Meunier and Morel theory, the MP is conceivable in the {\it restricted} range $\varepsilon>\varepsilon_{\rm m}(\db,A)$, and the so called {\it pseudo-spinodal point}, $\beta^*(A)$, at which $\varepsilon_{\rm m}$ reaches the inflexion point $\varepsilon^*(A)$ (the vertical line in the figure) signals the endpoint of metastability (the figure indicates that $\db^*(256^2)$ is somewhere in between 0.0038 and 0.0048). For $\db<\db^*$,  the agreement between theory and numerics is good. We attribute the progressive discrepancies for large $|\varepsilon|$s to a statistical underestimation for low probabilities and to a finite-size effect, presumably consequence of the continuum description of Eq. (\ref{eq:phi}). On the other hand, for $\db>\db^*$ the difference between theory and numerics becomes essential: while the theory predicts absence of metastability (no convex $P_{\db,A}$), we nevertheless still observe stationary states with convex $P$.  Moreover, as illustrated in the inset, the EPD curves for different values of  $\db>\db^*$ are not derivable one from another by re-weighting (as for $\db<\db^*$ they do, see Eq. (\ref{eq:reweighting})). This is an evidence of the fact that these stationary states do not obey Maxwell-Boltzmann statistics. An interesting question is whether the lifetime of the non-equilibrium stationary states for $\db>\db^*$ remains finite in the large-$A$ limit. This point will be discussed below. \\ 
\indent
At the pseudo-spinodal point $\db^*(A)$, the slope of the EPD at $\varepsilon^*(A)$ vanishes. We have estimated it by extrapolating in $\db$ our data for $(\partial_\varepsilon|_{\varepsilon^*} \ln P)_{\db,A}$ down to zero, obtaining values of $\db^*(A)$ in reasonable (better the larger the size) agreement with those of the DT (see Fig. \ref{fig:regimes}). The data approximately exhibit the size dependence $\db^*(A)  = C + D\, A^{-1/3}$, with a small $C$. The  $\db^*\sim A^{-1/3}$ scaling is compatible with the  DT, as we find for $\db\searrow 0$. 
The numerical results for $\db<\db^*$ are, hence, compatible with the DT picture, which predicts absence of equilibrium metastability for large $A$.  \\
A direct comparison with the theory is possible also for the lifetime $\tau(\db,A)$. {\it We assume} that the lifetime of the MP is the time needed to create a fluctuation with energy lower than the limit of stability $\varepsilon_{\rm m}(\db,A)$, i.e., $\tau/\tau_{\rm r}\sim P_{\db,A}(\varepsilon_{\rm m}(\db,A))^{-1}$. 
%
$\tau$ is to be measured in units of the  self-correlation time $\tau_{\rm r}$, since the  probability of a critical configuration is inversely proportional to the average number  of {\it uncorrelated} configurations before leaving the plateau. Using $P_{\db,A}(\varepsilon_{\rm m})$  from the solution of Eq. (\ref{eq:Pe}) (and neglecting the $\db$ dependence of $\tau_{\rm r}$) we have estimated the $\beta$-dependence of  $\tau$, which agrees with our numerical results in the DT validity region $\db<\db^*(A)$. The comparison is illustrated in Fig. \ref{fig:tau}, in which we also report the analytical expression for $\ln\tau$, that we have obtained in saddle-point approximation to Eq. (\ref{eq:Pe}), valid for $\db \searrow 0$ \cite{Berganza2014}:

\barray
\ln \tau(\db,A)\sim p\ln \frac{2w_q}{3\db} + \frac{4w_q^3}{27}\frac{1}{\db^2} + {\cal K}_A  
,
\label{eq:tauvsdb}
\earray
with $p=7$, and ${\cal K}$ being a decreasing function of the area, constant in $\db$. 
Interestingly,  assuming $\tau\sim I^{-1}$ with $I$ given by the Langer relation $I\sim {\rm Im}\,\phi(z\nearrow 0)$, leads alternatively for $\tau$ to the same law as in Eq. (\ref{eq:tauvsdb}) with $p=5$ (see Fig. \ref{fig:tau}), although this simpler approach does not provide information about ${\cal K}_A$. For fixed temperatures, the numerical data for $\tau$ systematically decrease with $A$, in agreement with the DT. \\ 
\indent
From our data of $\tau$ we estimate  the {\it dynamical temperature} $\dbn(A,t)$ at which the average lifetime is $t$. 
Remarkably, also our results for $\dbn$ present a clear scaling:

\beq
\dbn(A,t) = C_t\, A^{-1/3} + D_t
\label{eq:Ascaling}
\eeq
with nonzero  $D_t$ (see Fig. \ref{fig:regimes}). We define in this way a temperature endpoint of the stationary phase, defined by the temperature at which $\tau$ becomes small. Choosing   $t_{\rm min}=6\,10^3$ MC steps, we obtain a threshold, $\dbn(A,t_{\rm min})$, below which the stationary states are no longer observable in practice.  Eq. (\ref{eq:Ascaling})  suggests that in the large-$A$ limit there may be a nonzero temperature interval $[\bc:\bc+D_{t_{\rm min}}]$ in which the non-Boltzmann stationary states indeed survive with a nonzero lifetime. \\
\indent
Our findings are summarized in Fig. \ref{fig:regimes}. 
For $\db<\db^*(A)$ the local dynamics leads to Boltzmann DT-describable metastable states. On the other hand, for $\db>\db^*$ and sufficiently low, we expect stationary states with  non-Boltzmann statistics. 
Finally, for $\db \gtrsim \dbn(A,t_{\rm min})$, no stationary states, just off-equilibrium relaxation towards the ordered phase is observed.\\ 
\indent
The shrinking of the metastable interval may not be in contradiction with the results of the aforementioned pseudo-critical studies: the limit $\bs$, where $\tau_{\rm r}$ is supposed to diverge, cannot be reached in a dynamical scheme, in which metastability is supposed to end at  lower $\beta$ when $\tau_{\rm r}$ becomes of the order of $\tau$. The argument $\tau(\db^*,A)\sim \tau_{\rm r}(\db^*,A)$ is indeed compatible with the decreasing of $\db^*(A)$, since as $A$ increases, $\tau$ decreases and $\tau_{\rm r}$ anomalously increases \cite{Berganza2014}. This dynamical meaning of $\db^*$ is also compatible with our data, as we anticipate in Fig. \ref{fig:tau}.  \\
\indent
An important point  concerns the microscopic origin of these different behaviors, which remains unknown. In \cite{Nogawa2011Static} it is conjectured that the canonical spinodal point is associated to the Evaporation-Condensation transition \cite{Binder1980Critical,Binder2003Theory}, a finite-size effect  occurring at coexistence in the micro-canonical ensemble, and that has been observed for the PM in 2D (when the micro-canonical endpoint of the stable energy branch is again at a value $\db\sim A^{-1/3}$) \cite{Nogawa2011Evaporation,Nogawa2011Static,Janke1998Canonical} and 3D \cite{Bazavov2008Phase}.  The  authors of \cite{Nogawa2011Static} also conjecture the existence of a length scale $\ell$ (suggested to be the typical distance between critical droplets, $R_0$) 
such that the condition $A\sim\ell^2$ would trigger a crossover between different dynamical regimes (see also \cite{Bazavov2008Phase}). 
In any case, in the so called {\it deterministic region},  $A\gg R_0^2$, finite-size effects should disappear, as happens in the ILGM \cite{Rikvold1994Metastable}. The decreasing of $\tau$ with $A$ for arbitrarily large $A$ is essentially different with respect to the  ILGM paradigm, and implies a size-dependent nucleation rate. In \cite{Berganza2007}, it is proposed that, while the bulk term in $f_a$ of the ILGM is size-independent, in the PM case it may come from an entropy-maximizing constraint, such that clusters of a given color are confined to avoid the breaking of the symmetry between colors. In larger systems, clusters would be less confined since they contribute less to the magnetization. \\
\indent
The DT cannot validate this idea since it is obtained from an infinite-volume free energy, not  allowing for a microscopic formulation in terms of droplets. We are investigating an alternative scheme by estimating the form of the size-dependent droplet free-energy $f_{a,A}$.  The  dependence of the results on lattice, algorithm dynamics and $q$ value are also being examined. \\
\indent
Summing up, we have provided a picture of the metastable dynamics of the 2D PM. The MPs present a strong finite-size scaling, well described by the DT. Such an anomalous finite-size scaling, whose microscopic origin is controversial, is different with respect to the metastability in the ILGM, and may be found in other first-order transitions.\\
{\it Acknowledgments.} We acknowledge Kurt Binder, Ezequiel E. Ferrero and Per Arne Rikvold for critical comments on the manuscript. We gratefully acknowledge the support of NVIDIA Corporation with the donation of the Tesla K20 GPU used for this research.  

\bibliography{berganza-potts}
\bibliographystyle{apsrev4-1}

\end{document}